\documentstyle[aps,eqsecnum,preprint,umd]{revtex}
\def\be{\begin{eqnarray}}
\def\ee{\end{eqnarray}}
\def\MeV{\nobreak\,\mbox{MeV}}
\def\GeV{\nobreak\,\mbox{GeV}}
\def\fm{\nobreak\,\mbox{fm}}
\def\log{\nobreak\,\mbox{ln}}
\def\qbar{\overline{q}}
\def\ubar{\overline{u}}
\def\dbar{\overline{d}}
\def\sbar{\overline{s}}
\def\nubar{\overline{\nu}}
\def\bra#1{\langle #1|}
\def\ket#1{| #1\rangle}
\def\nucbra{\bra{ p}}
\def\nucbrap{\bra{ p'}}
\def\nucket{\ket{ p}}

\begin{document}
\draft

\preprint{\vbox{Submitted to Phys.\ Rev.\ C \hfill DOE/ER/40762--030\\
                \null\hfill UMPP \#94--097, hep-ph/9408326 \\ }}

\title{ Stranger in the Light: The Strange Vector Form Factors \\
of the Nucleon}

%

\author{Hilmar Forkel\thanks{Address after Nov. 1, 1994: European
Centre for Theoretical Studies in Nuclear Physics and Related
Areas, Villa Tambosi, Strada delle Tabarelle 286, I-38050
Villazzano, Italy}, Marina Nielsen%
\thanks{Permanent address: Instituto de F\'{\i}sica,
Universidade de S\~ao Paulo, 01498 - SP- Brazil.}, Xuemin
Jin and
Thomas~D. Cohen}
%


\address{Department of Physics and Center for Theoretical Physics\\
University of Maryland, College Park, Maryland 20742}
%
%

\maketitle

\vskip -0.5cm

\begin{abstract}

\vskip -0.5cm

A model combining vector meson dominance, $\omega - \phi$ mixing
and a kaon cloud contribution is used to estimate the strangeness
vector form factors of the nucleon in the momentum range of the
planned measurements at MIT-Bates and CEBAF. We compare our results
with some other theoretical estimates and discuss the nucleon
strangeness radius in models based on dynamically dressed, extended
constituent quarks. For a quantitative estimate in the latter
framework we use the Nambu-Jona-Lasinio model.

\end{abstract}
\pacs{PACS numbers: 14.20.Dh, 12.40Vv, 12.15Mm, 12.40.-y,
12.39.Ki }

\section{Introduction}

\label{intro}

Strangely, not all static low-energy properties of the nucleon
are yet measured. In particular, practically no experimental
information exists on the nucleon form factors of the strange-quark
vector current, the subject of our paper. One might ask why these
observables have so far escaped experimental scrutiny. Certainly
they are not easy to measure, requiring high-statistics neutral
current experiments at the limits of present capabilities, but
perhaps more importantly they were widely expected to be zero or
at least small.

In the absence of direct insight from QCD at low energies, these
expectations had been guided by hadron models, and most prominently
by the nonrelativistic quark model (NRQM) \cite{der75}. The latter
gained particular credibility by describing the low-lying meson
and baryon spectra better than other, and often more complex,
hadron models. However, this model is based either on the complete
neglect or at least on a drastic simplification of relativistic
effects, which seems hard to justify in the light quark sector.

This limited treatment does of course not only affect the valence
quarks. More importantly in our context, it excludes vacuum
fluctuations and, in particular, the existence of virtual
quark-antiquark pairs in the hadron wave functions. The nucleon,
for example, is described as consisting solely of up and down
constituent quarks, and the NRQM thus predicts the absence of
any strangeness distribution.\footnote{In some generalized
constituent quark models, the quarks have an extended structure
which can contain strangeness (see Sec. \ref{cqmsection}). }

It therefore came as a surprise to many when deep inelastic
$\mu -p$ scattering data from the European Muon Collaboration
(EMC) \cite{ash1} indicated a rather large strange quark contribution
to the singlet axial current of the nucleon. The EMC measured the
polarized proton structure function $g_1^p(x)$ in a large range
of the Bjorken variable, $x \in [0.01,0.7]$ \cite{ash1,ash2} and
found, after Regge extrapolation to $x=0$ and combination with
earlier SLAC data,
\be
\int_0^1 dx \, g_1^p(x) = 0.126\pm 0.010 \, (stat) \pm 0.015 \,
(syst) \;
\ee
at $Q^2 = 10 \GeV^2 / c^2$. Ellis and Jaffe's prediction \cite{el},
which neglected strange quark contributions, was significantly
larger: $0.175\pm 0.018$. From the above data, however, one
extracts a nonvanishing strange quark contribution $\Delta s =
-0.16\pm 0.008$ to the proton spin, or equivalently, via the
Bjorken sum rule, a substantial strangeness contribution to the
proton matrix element of the isoscalar axial-vector current.

The low-energy elastic $\nu -p$ scattering experiment E734 at
Brookhaven \cite{ahr87} complemented the EMC data by measuring
the same matrix element at smaller momenta,  $(0.4 \GeV^2 < Q^2
< 1.1 \GeV^2)$. The results of this experiment also contain
information on the strange vector form factors and will therefore
be discussed in more detail later, in Sec. \ref{exptsection}.
The extracted axial vector current form factors are consistent
with the muon scattering data from the EMC \cite{kap88}.

The scalar channel provides additional, although indirect,
experimental evidence for a significant strangeness content of
the nucleon \cite{che,das,don,ga}. It stems from the nucleon
sigma term, $\Sigma_{\pi N}$, which can be extracted from
pion-nucleon scattering data. Defining the ratio $R_s$ as
\be
R_s = {\nucbra \sbar s\nucket\over\nucbra \ubar u + \dbar d
+ \sbar s\nucket} \; , \label{r}
\ee
($u$, $d$ and $s$ are the up, down and strange quark fields,
and $\nucket$ denotes the nucleon state) one finds $R_s\simeq
0.1 - \, 0.2$. Again, the resulting $\nucbra \sbar s\nucket$
matrix element is surprisingly large, of up to half the magnitude
of the corresponding up-quark matrix element. This implies that
the nucleon mass would be reduced by $\approx 300 \MeV$ in a
world with massless strange quarks. The analysis leading to the
quoted range of  $R_s$ values is, however, not uncontroversial,
and smaller values  have been suggested \cite{bla88}.

In any case, there is evidence that strange quarks play a more
significant role in the nucleon than the quark model suggests.
One direction for the further exploration of this intriguing
issue is to study the nucleon matrix elements of strange-quark
operators in other channels. Since the electroweak neutral currents
provide us with experimental probes in the vector channel, the
strange vector current has recently received considerable
attention. In particular, a number of devoted experiments at
MIT-Bates \cite{mit} and CEBAF \cite{ce17,ce4,ce10} (see Sec.
\ref{exptsection}) will measure part of the corresponding form
factors at low momenta.

In this situation it would clearly be desirable to have reliable
theoretical predictions for these form factors. Unfortunately,
the present status of our theoretical understanding is qualitative
at best. A phenomenological estimate and some model calculations
have been performed, but all results have large theoretical
uncertainties and are partially inconsistent with each other.

Since the sea quark distribution in hadrons arises from a subtle
interplay of quantum effects in QCD, their reproduction in hadron
models is much more challenging than the calculation of the
standard static observables. Lattice calculations of strange
quark distributions have not yet been performed since the
disconnected quark loops increase the computational demands of
form factor calculations substantially.

The particular value of the strange quark mass, which is neither
light nor heavy compared to the QCD scale $\Lambda$, additionally
complicates the theoretical situation. In contrast to the light
up and down quarks, the effects of the heavier strange quark
are much harder to approach from the chiral limit, {\it i.e.}
by an expansion in the quark mass. On the other hand, the
strange quark is too light for the methods of the heavy-quark
sector, {\it e.g.}  the nonrelativistic approximation or the
heavy-quark symmetry, to work.

The purpose of this paper is to assess the present theoretical
status by discussing some of the proposed models, and to extend
our previous estimate of the nucleon strangeness radius in a
vector dominance model to the calculation of the strangeness
vector form factors in the momentum region of interest for the
planned experiments.

In Sec. \ref{2section} we will set up our
notation, give the basic definitions and discuss the presently
available and soon to be expected experimental information.
Section \ref{3section} contains a review of some theoretical
estimates for the strangeness radius and magnetic moment,
including the pole fit of Jaffe  \cite{ja}, a Skyrme model
calculation \cite{par91}, the kaon cloud model estimate
\cite{mus93} and our own estimate based on vector meson
dominance and $\omega-\phi$ mixing \cite{nos}. We conclude
this section with a new calculation in the framework of
extended constituent quark models, where the light quarks
are dressed with a quark-antiquark cloud and carry strangeness.
The fourth section describes the calculation of the strange
vector form factors in our model, discusses our results and
compares them to the pole fit estimate and the sometimes used
dipole form. In Sec. \ref{5section} we summarize the paper
and present our conclusions.

\section{Basic concepts and experimental status }
\label{2section}

In order to quantify the concept of the nucleon's strangeness
distributions in the language of QCD, one considers nucleon
matrix elements of the form
\be
\nucbra \, \sbar \, \Gamma s \, \nucket,
\ee
where $\Gamma$ stands for a Dirac matrix which selects the
space-time quantum numbers of the matrix element. These matrix
elements are in general scale--dependent. As indicated in the
introduction, we will discuss in particular the matrix element
of the strange vector current ({\it i.e.}  $\Gamma = \gamma_\mu$),
which, due to strangeness conservation, is independent of the
subtraction point of the current operator.

Most of the theoretical work has so far focused on the lowest
nonvanishing radial moments of the strangeness spin and charge
distributions. These are the strangeness magnetic moment $\mu_s$,
defined as the z-component of
\be
\vec{\mu}_s = \frac12 \int d^3 r \,\nucbra \vec{r} \times \sbar
\vec{\gamma} s \nucket,
\ee
and the square of the strangeness radius,
\be
r^2_s =   \int d^3 r \nucbra  \vec{r}^{\,\, 2} \,\, \sbar
\gamma_0 s  \nucket.
\ee

In analogy to the electromagnetic case, the momentum dependence of
the strangeness current matrix elements is contained in two form
factors,
\be
\nucbrap  \sbar \gamma_\mu s \nucket = \overline{N}(p^\prime)
\left(\gamma_\mu F_{1}^s(q^2) +
i\frac{\sigma_{\mu\nu}q^\nu} {2M_N} F_{2}^s(q^2)
\right) N(p), \label{sff}
\ee
where $q = p^\prime - p$ and $N$ is the free Dirac spinor of the
nucleon. Strangeness conservation and the zero overall strangeness
charge of the proton imply $F_{1}^s(0) = 0$.

It is often convenient to use the electric and magnetic form
factors introduced by Sachs
\be
G_E^s(q^2) &=& F_1^s(q^2) + {q^2\over4M_N^2}F_2^s(q^2),
\nonumber\\*[7.2pt] G_M^s(q^2) &=& F_1^s(q^2) + F_2^s(q^2) ,
\label{sachs}
\ee
which describe the strangeness charge and current distribution,
respectively. The above matrix elements for $\mu_s$ and $r^2_s$
are simply related to the form factors by
\be
\mu_s = G_M^s(0), \qquad \quad r^2_{s} =  6 \frac{d }{d q^2}
G_E^s(q^2) |_{q^2=0}. \label{mur}
\ee
Note that we use the Sachs form factor, which corresponds to the
physical charge distribution, to define the strangeness radius.
An alternative definition, based on the Dirac form factor
({\it i.e.} $F_{1}^s$), has also been used in the literature.

For the following discussion it will be useful to recall the
relation of the strangeness vector current to the other neutral
vector currents and to the standard $U(3)$ current nonet in the
three-flavor sector,
\be
J^{a}_\mu = \qbar \gamma_\mu \frac{\lambda^a}{2} q, \label{flavcurr}
\ee
with $\lambda^0 \equiv \sqrt{2/3}$ and ${\rm tr} \{\lambda^a
\lambda^b \} = 2 \delta^{a b}$. In the standard flavor basis
defined by Eq. (\ref{flavcurr}), the neutral currents and the
corresponding operators for the electric charge, the baryon number
and the hypercharge have the form
\be
J^{el}_\mu  &=& \qbar \gamma_\mu \, Q^{el} q = J^{3}_\mu +
\frac{1}{\sqrt{3}} \, J^{8}_\mu, \,\,\,\,\,\qquad  Q^{el} = T^3 +
\frac{Y}{2} = \frac{\lambda^3}{2} + \frac{\lambda^8}{2 \sqrt{3}} ,\\
J^{B}_\mu &=& \qbar \gamma_\mu B q =  \frac{1}{\sqrt{6}} \, J^{0}_\mu,
\qquad \qquad  \qquad B = \frac13 , \\
J^{Y}_\mu &=& \qbar \gamma_\mu Y q = \frac{2}{\sqrt{3}} \, J^{8}_\mu,
 \qquad \qquad  \qquad  Y = B - S = \frac{\lambda^8}{ \sqrt{3}},
\label{hyp}
\ee
and the strangeness current can be written as
\be
J^{s}_\mu &=&  \sbar \gamma_\mu s = J^{B}_\mu - \, J^{Y}_\mu.
\label{strcurr}
\ee
Note that we use the non-standard sign convention of Jaffe \cite{ja}
for the strangeness current. This leads to the negative sign of the
strangeness contribution to the hypercharge in Eq. (\ref{hyp}).

\subsection{Experimental information}
\label{exptsection}

No direct measurement of the strange vector form factors has
yet been performed. However, there is some (although indirect and
rather uncertain) experimental evidence for them to be nonvanishing.
A recent reanalysis \cite{gar93} of the E734 experiment at BNL
\cite{ahr87}, which measured the elastic $\nu p$ and $\nubar p$
differential cross sections in the momentum range $0.2 \GeV^2
< Q^2 < 1.2 \GeV^2 $ to constrain the strange axial-vector form
factor of the proton, indeed mildly favors nonzero vector form
factors.

As suggested by Kaplan and Manohar \cite{kap88}, the authors of
Ref. \cite{gar93}
refitted the E734 cross sections and, in contrast to the original
analysis, allowed the strange vector form factors to be different
from zero. Their analysis also takes the neutron electromagnetic
form factor into account and fits the value of the mass parameter
in the axial form factors, $M_A$. The observed elastic $\nu p$
and $\nubar p$ scattering events and their statistical and
systematical errors were binned in only seven $Q^2$ regions, which
limits the number of fit parameters which can be determined from
this data set. Garvey {\it et al.} \cite{gar93} therefore determine
only the lowest nonvanishing moments of the strange axial [$G^s(Q^2$)]
and vector [$F^s_1(Q^2), F^s_2(Q^2)$] form factors, and assume the
same momentum dependence as in the corresponding electromagnetic
form factors.

Since no symmetry arguments connect the $SU(3)$ singlet form
factors with their octet counterparts, this is clearly a rather
strong assumption. As we will discuss below, it is neither supported
by the existing theoretical estimates nor by simple ``quark-core
plus meson-cloud'' pictures of the nucleon. In the latter, one
expects
the electromagnetic charge and current distribution to be carried
by a valence quark core and a meson cloud. Due to the absence of
strange valence quarks one would expect a different, more outwards
shifted strangeness distribution in this picture. We will discuss
the dipole form further in the last section of this paper, where we
compare it with our results and with another phenomenological
estimate.

We do not, however, expect a different momentum dependence of
$F^s_1$ and  $F^s_2$ to change the fits of Ref. \cite{gar93}
significantly since the neutrino cross sections are
more sensitive to the strange axial form factor. The strong
correlation between the fit values of $G^s(0)$ and $M_A$ noted
in Ref. \cite{gar93} might, however, be enhanced by the assumed
dipole form of $G^s$ with  $M^s_A = M_A$.

Although the existing data are not sufficient to strongly constrain
any of the strange form factors, the difference in the  $Q^2$
behavior of the neutrino -- and antineutrino -- proton cross
sections seems to favor a finite, negative strangeness radius
and a negative strangeness magnetic moment. As we will discuss
in Sec. \ref{kclsection}, these signs are expected if the nucleon's
strangeness distribution arises mainly from fluctuations of its
wave function into the hyperon plus kaon configuration.

Clearly, much more stringent experimental constraints on the
strangeness form factors would be desirable. Some time ago,
McKeown \cite{mck89} and Beck \cite{bec89} pointed out that
neutral current form factors could be measured in parity-violating
electron scattering experiments. Four such experiments are at
present in different stages of preparation and will soon provide
the first direct measurements of sea quark effects in low-energy
observables. SAMPLE at MIT-Bates plans to measure the value of
$F^s_2$ at $Q^2 = 0.1 \GeV^2$ and will start to take data in
less than a year. Three further experiments are located at CEBAF
and will measure different combinations of the strange form
factors in a larger range of momentum transfers. We refer the
reader to Ref.  \cite{fatreview} for a review of these experimental
programs.

The sensitivity of this first generation of parity-violating
elastic electron-proton scattering experiments will be high
enough to distinguish among some of the present theoretical
estimates of the strangeness radius and can therefore provide
valuable and much needed constraints for nucleon models.

\section{The vector strangeness radius and magnetic moment }
\label{3section}

In this section we will discuss some of the theoretical estimates
of the strangeness radius and magnetic moment. We will also add
a new calculation of these quantities in models containing extended
constituent quarks. The evaluation and discussion of the form factors
for a larger momentum transfer region will be the subject of the
subsequent section.

Let us point out from the beginning that all of the existing
estimates are strongly model dependent and that none of them should
be considered as a firm and reliable prediction. However, they give
at least insight into the order of magnitude of the form factors and
can thus provide motivation and guidance for future experiments. And,
equally important, they explore and test different physical mechanisms
for the appearance of a strangeness content in the nucleon.

\subsection{The pole fit model}
\label{polefit}

The first estimate of the strangeness radius and magnetic moment
by Jaffe \cite{ja} is based on H{\"o}hler {\it et al.}'s pole fit
of the isoscalar electromagnetic form factor of the nucleon
\cite{hoe76}, which assumes these form factors to be dominated by
three (zero-width) isoscalar vector ($ I^G \, J^{PC} = 0^- 1^{- -}$)
meson states: the physical $\omega(780)$ and $\phi(1020)$ mesons
and a third, higher-lying pole, which is intended to summarize the
high-mass resonance and continuum contributions.  (For a simpler
estimate on the basis of a one-pole model and $\omega - \phi$ mixing
see ref. \cite{par91}.)

Adopting  H{\"o}hler's three-pole ansatz also for the strangeness
form factors, Jaffe writes $F_1^s$ in the form of a once--subtracted
dispersion relation [making use of $F_1^s(0) = 0$],
\be
F_1^s(q^2) = \sum_{i=1}^3 \, a_i^{(s)} \, \frac{q^2}{m^2_i - q^2} ,
\label{f1j}
\ee
whereas $F_2^s$ is not normalized and taken unsubtracted,
\be
F_2^s(q^2) = \sum_{i=1}^3 \, b_i^{(s)} \, \frac{1}{m^2_i - q^2}.
\label{f2j}
\ee
H{\"o}hler's form factors $F_{1,2}^{(I=0)}$ for the isoscalar part
of the electromagnetic current ({\it i.e.}, $J^{(I=0)}_\mu = 1/2
J^{Y}_\mu$) have exactly the same form. The corresponding masses
$m_i$ and coupling parameters $a_i^{(I=0)}, b_i^{(I=0)}$ have been
determined by fits to the experimental data (see Table I).

Jaffe fixes all three masses in Eqs. (\ref{f1j})) and (\ref{f2j})
-- including the one of the third pole -- at the same values as
in the isoscalar form factors of Ref. \cite{hoe76}, see Table I.
The couplings $a_{1,2}^{(s)}$ and $b_{1,2}^{(s)}$ can be related
to the corresponding couplings $a_{1,2}^{(I=0)}$ and $b_{1,2}^{(I=0)}$
by a simple quark counting prescription, starting from the flavor
wave functions of the $\omega$ and $\phi$ mesons,
\be
|\omega> &=& \cos\epsilon\,\frac{1}{\sqrt{2}} \left( |\ubar\gamma_\mu
u> + |\dbar\gamma_\mu d> \right) - \sin\epsilon\,| \sbar \gamma_\mu
s > , \nonumber \\*[7.2pt]
|\phi> &=& \sin\epsilon\,|\omega_0> + \cos\epsilon\,|\phi_0>\ .
\label{states}
\ee
The small angle $\epsilon=0.053$ \cite{jain} parametrizes the
deviation from the ideally mixed states. The current-meson
couplings are obtained from the assumption that a quark $q_i$
with flavor $i$ in the vector mesons couples only to the current
$\qbar_i \gamma_\mu q_i$ of the same flavor, and with
flavor-independent strength $\kappa$:
\be
g(\omega,J^{I=0}) &=& \frac{\kappa}{\sqrt{6}} \sin (\theta_0 +
\epsilon), \qquad \,\,
g(\phi,J^{I=0}) = - \frac{\kappa}{\sqrt{6}} \cos (\theta_0 +
\epsilon), \nonumber \\
g(\omega,J^{s}) &=& - \kappa \sin \epsilon, \qquad \qquad \qquad
g(\phi,J^{s}) = \kappa \cos \epsilon,
\ee
where $\theta_0$ is the ``magic angle'' with $\sin^2 \theta_0 =
1/3$. The vector-meson nucleon couplings are parametrized as
$g_i (\omega_0,N) = g_i \cos \eta_i ,\, g_i(\phi_0,N) =  g_i
\sin \eta_i$, where $i = 1,2$ denotes the $\gamma_\mu$ and
$\sigma_{\mu \nu} q^\nu$ couplings.

Comparing the above parametrization of the couplings with
H{\"o}hler's fitted values for $a_{1,2}^{(I=0)}$ and
$b_{1,2}^{(I=0)}$
leads to phenomenological values for the $\eta_i$ and $\kappa
g_i$. The strangeness current couplings $a_{1,2}^{(s)}$ and
$b_{1,2}^{(s)}$ are then obtained by simply replacing $J^{I=0}$
with  $J^{s}$. The remaining couplings  $a_{3}^{(s)}$ and
$b_{3}^{(s)}$ are determined by imposing the (rather mild)
asymptotic constraints $\lim_{q^2 \rightarrow \infty}  \,\,
F_1^s(q^2) \rightarrow 0, \quad  \lim_{q^2 \rightarrow \infty}
\,\, q^2 \, F_2^s(q^2) \rightarrow 0$.

The strangeness radius and magnetic moment can now be obtained
from Eqs. (\ref{sachs}) and (\ref{mur}). Taking the average
of the three fits in Ref. \cite{hoe76}, Jaffe finds $r_s^2 =
(0.14 \pm 0.07) \, {\rm fm}^2$ and $\mu_s = - (0.31 \pm 0.09)$.
Note that these estimates are rather large -- of the
order of the corresponding electromagnetic moments of the neutron --
and that they do not rely on a specific nucleon model. However,
bias is introduced through the assumed three-pole ansatz for the
form factors and the identification of the two light poles with the
physical $\omega$ and $\phi$ mesons. Furthermore, the final form of
the ansatz relies on the conditions for the asymptotic behavior
and on the parametrization of the couplings (which works well
in the electromagnetic case).

As pointed out by Jaffe, the pole fit results depend crucially on
H{\"o}hler's  identification of the second pole with the physical
$\phi$ meson, with its large strange quark content and its
surprisingly strong, OZI violating coupling to the nucleon
\cite{gen76}. It should further be noted that adopting the
third pole mass from  H{\"o}hler {\it et al.} is hard to justify.
The third pole does not correspond to a well-defined state, but
rather summarizes unknown higher-lying resonance and continuum
contributions, which will likely have a different distribution of
strength for the hypercharge and strange currents.

Let us finally mention that the sign of $r^2_s$ is opposite to
the one suggested by Garvey {\it et al.}'s reanalysis of the BNL
neutrino scattering data and found in other models, and that the
three-pole ansatz does not allow the form factors to decay with
the large powers of $q^2$ established via quark counting rules.

\subsection{Skyrme model estimate}

The first nucleon model estimate of the low-momentum strangeness
form factors was based on the Skyrme model \cite{par91}. The latter
is a topological soliton model, built on a chiral meson lagrangian.
Many variants of this model and of its treatment exist and are
described {\it e.g.} in the reviews \cite{skyrrep}. The extension
to the $SU(3)$ flavor sector and its breaking is not unique and
allows an even wider choice in the specific approach, which of
course leads to some ambiguity in the results.

The authors of Ref. \cite{par91} choose the original Skyrme model
\cite{sky58} and introduce flavor symmetry breaking by non-minimal
derivative terms in the lagrangian. After canonical quantization
in the restricted Hilbert space of collective and radial
excitations, the Hamilton operator is diagonalized by treating the
symmetry breaking terms exactly \cite{yab88}. This approach in
some sense combines the two conventional quantization schemes of
the $SU(3)$ skyrmion. From the strange vector form factors  Ref.
\cite{par91} obtains  $r_s^2 = - 0.10 \,{\rm fm}^2$ and $\mu_s =
-0.13$. In an extended Skyrme model containing vector mesons,
these values drop by about a factor of two in magnitude, and
the sign of the strangeness radius changes \cite{par92}.

It is interesting that these results for nucleon sea quark
distributions come from a model which describes quark physics
rather indirectly in terms of meson fields. However, considerable
theoretical uncertainties are associated with the Skyrme model
estimates of the strange form factors. Besides the already
mentioned ambiguities in the specific choice of lagrangian and
$SU(3)$ extension, there are several additional problems.

One serious concern is that the calculation of strange matrix
elements in Skyrme-like models ultimately involves the small
difference of two large but uncertain quantities, and hence is
intrinsically unreliable. The source of this difficulty is that
the strange vector current, which in QCD is understood at the quark
level, must be described in terms of the Noether and topological
currents in the Skyrme picture.

The strange current, in particular, is obtained from the relation
of baryon number and hypercharge currents, Eq. (\ref{strcurr}):
\be
J_{\mu}^s = J^B_{\mu} - J^Y_{\mu} \; . \label{sby}
\ee
Thus,  $ r^2_s$ is given by  $r^2_B  -  r^2_Y$. However, both
$ r^2_B$  and $ r^2_Y$ are of order 1 ${\rm fm}^2$ so that their
difference is almost an order of magnitude smaller than the two
separately. The above--mentioned problem arises since quantities
calculated in the
Skyrme model have a typical accuracy at the 30\% level (without
introducing large numbers of parameters), which might even be
expected from a model justified to leading order in a $1/N_c$
expansion, with $N_c = 3$.

Therefore, one concludes that the Skyrme model
cannot accurately determine the strangeness radius to within
several hundred percent, unless there are special reasons to
believe that the errors in  $r^2_B$  and $r^2_Y$ are strongly
correlated.  However, it is hard to argue for correlations in
the errors, since they come from very different parts of the
model. The baryon current is of topological origin, and it is
difficult to understand why it should ``know'' about errors in
the hypercharge current, which is a Noether current.

There is also a more fundamental concern with the Skyrme model
calculation. The Skyrme model, and the approximations used to
treat it, are generally believed to be justified in the large-$N_c$
limit of QCD; effects of subleading contributions in $1/N_c$
are probably not reliable. This casts serious doubts on whether
the Skyrme model (or any other large-$N_c$ model) can ever make
contact with strange matrix elements. After all, in a conventional
treatment of the nucleon any strange quark matrix element must be
ascribed to a Zweig-rule violating amplitude. But, as argued by
Witten \cite{wit79} many years ago, Zweig's rule is exact in the
large-$N_c$ limit.

Apart from the question of whether the Skyrme model can give
reliable strange quark matrix elements in light of the need for
Zweig rule violations, there seems to be a paradox, since the
approach of Ref. \cite{par91} continues to give nonzero strange
matrix elements even as the parameters are pushed towards their
$N_c \rightarrow \infty$ values. How can the model give {\it any}
nonzero strange matrix element in the limit as $N_c \rightarrow
\infty$?

The resolution of this ``paradox'' is simple: Since the nucleon
is defined to be a member of the lowest $SU(3)_{\rm flavor}$ octet
with $N_c$ being any multiple of three, there are necessarily
valence strange
quarks in the nucleon for $N_c > 3$. To make a flavor octet out of
$N_c$ quarks, one groups the quarks in triples of flavor singlets
except for the last three, which are coupled to an octet. The key
point is that each of these groups of three quarks in a singlet
state
contains one strange quark.  Thus, a nucleon with $N_c > 3$ will
contain $\frac{N_c}{3} -1$ valence strange quarks.  A
consequence of this fact is that the standard relation  (\ref{sby})
between hypercharge, baryon number and strangeness must be modified
for $N_c \ne 3$:
\be
J_{\mu}^s = \frac{N_c}{3} J^B_{\mu} - J^Y_{\mu} \; . \label{sby2}
\ee
Given the net strangeness content of nucleons in the large $N_c$
world, there is no need for Zweig rule violating amplitudes.
Indeed, as $N_c \rightarrow
\infty$  the number of valence strange quarks diverges: the nucleon
becomes infinitely strange. This fact explains why the Skyrme model
can have nonzero strangeness matrix elements as $N_c \rightarrow
\infty$.

The resolution of this paradox does not, however, affect the general
conclusion that the ability of the Skyrme model to describe strange
quark matrix elements in the nucleon is likely to be fundamentally
limited, due to its large $N_c$ character. To describe our $N_c = 3$
world in the Skyrme model, one must start at the $N_c
\rightarrow \infty$ world and attempt to extrapolate back, with
inherent uncertainities in the $1/N_c$ correction terms. This
procedure may be quite sensible for some quantities, if the
underlying physics is essentially similar in the $N_c=3$ and
large $N_c$ worlds. It is likely to be problematic for strange
quark matrix elements, however,  since the character of the
physics completely changes as $N_c$ goes from infinity down to
three. At large $N_c$ there are a large number of valence strange
quarks and Zweig's rule is exact, while at $N_c=3$ there are no
valence strange quarks and the entire contribution is due to
Zweig rule violations.

\subsection{The kaon-cloud model}
\label{kclsection}

In order to complement the pole and Skyrme model calculations,
Musolf and Burkardt \cite{mus93} estimated the strange vector
matrix elements in yet another way, by considering the
contributions arising from a $K$ -- $\Lambda$ loop. The
heavier $K$ -- $\Sigma$ intermediate states have not been
taken into account. As pointed out in Ref. \cite{mus93}, the
corresponding couplings might, however, be significantly enhanced
beyond their $SU(3)$ values, which could make this contribution
relevant.

In contrast to earlier attempts \cite{hol90}, the calculation of
Ref. \cite{mus93} uses the phenomenological meson-baryon form
factors of the Bonn potential \cite{bonn} to cut off the loop
momentum and maintains gauge invariance via additional ``seagull''
vertices. The latter are generated from the Bonn form factors
by the minimal-substitution prescription. (For an alternative
choice of these form factors see Ref. \cite{koe94}.) The
interaction of the strange vector current with the hadrons in
the loop is treated as pointlike.

The above model is very likely to be too simple to provide
quantitative predictions, since {\it e.g.} contributions from
other but the lightest hadrons are not taken into account.
However, the motivation of the authors of Ref. \cite{mus93} was
not so much to end up with quantitative predictions, but rather
to explore and discuss qualitative features of loop contributions.

The strange magnetic moment obtained in this approach lies in
the range $\mu_s = -(0.31 - 0.40)$, dependent on the loop
cutoff value, and is of about the same magnitude as the pole and
Skyrme model predictions. The Sachs strangeness radius $r_s^2 =
- (0.027 - 0.032) \,{\rm fm}^2$, however, has the sign opposite
to the pole prediction, and its magnitude is smaller by a factor
of 3 to 5.

In the chiral limit, the strangeness radius develops an infrared
divergence from the meson propagator in the loop integral. It is
therefore very sensitive to the meson mass if the latter becomes
small. The Sachs radius, for example, would be an order of magnitude
larger if the pion would replace the kaon in the loop, as it does
in meson cloud models for the isovector part of the electromagnetic
form factors. The $SU(3)$ breaking effects are less pronounced for
the magnetic moment, where the corresponding loop integrals are
infrared finite.

Finally, in the absence of an agreement under the current
theoretical estimates for the sign of the strangeness radius,
another feature of the kaon cloud picture should be stressed: it
provides a simple and intuitive argument for the origin of the
sign of $r^2_s$. Since the (in our convention) negative strangeness
charge in the loop is carried by the kaon, which is less than half
as heavy as the lambda and thus reaches out further from the common
center of mass, it contributes dominantly to $r^2_s$ and determines
its negative sign.

This reasoning is analogous to the standard explanation for the
negative sign of the electromagnetic charge radius of the neutron
due to the negatively charged pion cloud. However, this static
picture is, of course, oversimplified and neglects, in particular,
recoil effects. The relevance of the latter can be seen in the
dependence of the strangeness radii on the involved mass scales.
The Dirac strangeness charge radius, for example, becomes positive
for pointlike kaons ({\it i.e.} for large values of the cutoff
mass in the meson-baryon form factors), whereas it stays negative
if the kaon mass is replaced by the pion mass \cite{mus93}.

\subsection{Kaon Cloud and Vector Meson Dominance}
\label{ourVMD}

In this section we review our model for the strangeness form
factors introduced in Ref. \cite{nos}. It combines an intrinsic
form factor, taken for definiteness from the kaon cloud model
discussed above, with vector meson dominance (VMD)  \cite{sa}
contributions and $\omega - \phi$ mixing. We are focusing in
the present section on the basic ideas underlying this approach
and will give more details, together with the calculation of the
full form factors, in Sec. IV.

The VMD hypothesis can, in its most general form, be summarized
in terms of current field identities (CFI) \cite{kro67}, which
state the proportionality of the electromagnetic current and
the field operators of light, neutral vector mesons with the
same quantum numbers. The isocalar CFI has thus the general
form
\be
J_\mu^{(I=0)} = A_\omega \, m_\omega^2 \, \omega_\mu + A_\phi
\, m_\phi^2 \, \phi_\mu, \label{cfi1}
\ee
with the couplings  $A_\omega$, $A_\phi$ yet to be fixed.
Generalizing the VMD hypothesis to the strangeness current, we
write an analogous CFI for $J^s$:
\be
J_\mu^s = B_\omega \, m_\omega^2 \, \omega_\mu + B_\phi \,
m_\phi^2 \, \phi_\mu. \label{cfi2}
\ee
It is convenient to combine these two CFI's into a vector equation,
so that the couplings form the elements of a matrix
\be
\hat{C}_{I=0,s} = \left(\begin{array}{cc}
A_\omega & A_\phi \\ B_\omega & B_\phi \end{array} \right).
\label{C1}
\ee
In order to fix the couplings in eq. (\ref{C1}), we sandwich the
CFI's between the physical vector meson states and the vacuum to
obtain the following representation for $\hat{C}$:
\be
\left(\begin{array}{cc} <0\,|\, J_\mu^{(I=0)} \,|\,\omega> &
<0\,|\, J_\mu^{(I=0)} \,|\,\phi>\\  <0\,|\, J_\mu^s \,|\,\omega>&
<0\,|\, J_\mu^s\, |\,\phi> \end{array} \right)\ = \varepsilon_\mu \,
\hat{C}_{I=0,s}  \left(\begin{array}{cc} m_\omega^2 & 0 \\
0 & m_\phi^2 \end{array} \right) . \label{C2}
\ee
($\varepsilon_\mu$ describes the polarization state of the vector
mesons) We now determine the matrix elements on the left-hand
side of eq. (\ref{C2}) from the simple $U(3)$ flavor counting rule
discussed in section \ref{polefit}. It states that the matrix
element of the quark vector current of flavor $i$ between the vacuum
and the flavor-$j$ component of the vector meson $V$ ($V$ stands for
$\omega$ or $\phi$) is diagonal in flavor and of universal strength
$\kappa$, {\it i.e.}
\be
 <0| \, \qbar_i \gamma_\mu  q_i \,|\, (\qbar_j q_j)_V\, > \, =
\kappa \, m^2_V \, \delta_{i j} \, \varepsilon_\mu. \label{me}
\ee
With the flavor wave functions of $\omega$ and $\phi$ from
section \ref{polefit} and the currents $J_\mu^{I=0} = 1/2
J_\mu^{Y}$  and $J_\mu^{s}$ given in eqs. (\ref{hyp}) and
(\ref{strcurr})) we then obtain
\be
\hat{C}_{I=0,s}(\epsilon) = \kappa \, \left(\begin{array}{cc}
{1\over\sqrt{6}}\sin(\theta_0 + \epsilon) &
{-1\over\sqrt{6}} \cos(\theta_0 + \epsilon)\\
-\sin\epsilon & \cos\epsilon
\end{array} \right)\ .
\label{C}
\ee
The same prescription for the couplings was used in Jaffe's pole
fit, see section \ref{polefit}. Both the magic angle $\theta_0 =
\tan^{-1} (1/\sqrt{2})$ and the mixing angle $\epsilon$, which
relates the ideally mixed flavor states of the vector mesons to
their physical states and is small and positive
\cite{rev}, have been introduced in Sec. \ref{polefit}. In the
following, we will use $\epsilon=0.053$, which has been determined
in Ref. \cite{jain} from the $ \phi \rightarrow \pi + \gamma $
decay width and  is consistent with the decay width of $\phi
\rightarrow \pi^+ +\pi^- +\pi^0$. Despite the small value of
$\epsilon$, the vector meson mixing will generate the dominant part
of our result for the strangeness radius \cite{nos}, which
consequently acquires a rather strong  $\epsilon$ dependence.

The CFI's lead to a general expression for the form factors. To
derive it, we first note that Eqs. (\ref{cfi1}) and (\ref{cfi2}),
together with the requirement of strangeness and hypercharge
conservation, imply $\partial^\mu V_\mu = 0$ ($V_\mu$ stands
for either $\omega_\mu$ or $\phi_\mu$), which simplifies the
field equations to
\be
( \Box + m^2_V ) \, V_\mu = J_\mu^{(V)} \label{feq}
\ee
and therefore also implies that the vector meson source currents
are conserved ($\partial^\mu J_\mu^{(V)} = 0$). We now take
nucleon matrix elements of the field equations (\ref{feq}) and
use the CFI's to write
\be
\left(\begin{array}{c} <N(p') | \, J_\mu^{(I=0)} \,| N(p)>
\\  < N(p')|\, J_\mu^s \,| N(p)> \end{array} \right)\ =
\hat{C}_{I=0,s}(\epsilon)  \left(\begin{array}{cc}
\frac{m^2_\omega}{m_\omega^2-q^2} & 0\\
0 & \frac{m_\phi^2}{m_\phi^2-q^2}
\end{array} \right)\
\left(\begin{array}{c} <N(p') |\, J_\mu^{(\omega)} \,| N(p)>
\\  < N(p')|\, J_\mu^{(\phi)} \,| N(p)> \end{array} \right).
\label{feq2}
\ee
Is is convenient to reexpress the vector meson source currents
in terms of currents with the same $SU(3)$ transformation
behavior as $J^{(I=0)}$ and $J^s$, which we will denote as
{\it intrinsic} ($J_{in}$). The corresponding transformation
can be written as
\be
\left(\begin{array}{c} J_\mu^{(\omega)}   \\
 J_\mu^{(\phi)} \end{array} \right) = \hat{D}_{I=0,s}
\left(\begin{array}{c} J_{in, \mu}^{(I=0) }    \\
 J_{in, \mu}^{(s)} \end{array} \right).
\ee
Applying it on the right-hand side of eq. (\ref{feq2}) and
separating the nucleon matrix elements into form factors,
according to eq. (\ref{sff}) and its analog for $J_\mu^{(I=0)}$,
we obtain our general VMD expression for the form factors:
\be
\left(\begin{array}{c}
F^{I=0}(q^2)\\
F^s(q^2)
\end{array} \right) = \hat{C}_{I=0,s}(\epsilon)
\left(\begin{array}{cc}
\frac{m^2_\omega}{m_\omega^2-q^2} & 0\\
0 & \frac{m_\phi^2}{m_\phi^2-q^2}
\end{array} \right)\hat{C}_{I=0,s}^{-1}(\epsilon)
\left(\begin{array}{c}
F_{in}^{I=0}(q^2)\\
F_{in}^s(q^2)
\end{array}\right)\ .
\label{form1}
\ee
According to their definition, the intrinsic form factors describe
the extended source current distribution of the nucleon to which the
vector mesons couple. Since both $J^{(I=0)}$, $J^{(s)}$ and their
intrinsic counterparts  $J_{in}^{(I=0)}$, $J_{in}^{(s)}$
are conserved, the full and the intrinsic form factors in eq.
(\ref{form1}) have the same normalization at $q^2 = 0$. This
immediately implies $\hat{D}_{I=0,s} = \hat{C}_{I=0,s}^{-1}$ and
has been anticipated in writing eq. (\ref{form1}). Combining Eqs.
(\ref{form1}) and (\ref{C}), we finally obtain
\be
\left(\begin{array}{c}
F^{I=0}(q^2)\\
F^s(q^2)
\end{array} \right) & = &
\left(\begin{array}{cc}
\frac{m^2_\omega}{m_\omega^2-q^2}\frac{\sin(\theta_0+\epsilon)
\cos\epsilon}{\sin\theta_0}-\frac{m^2_\phi}{m_\phi^2-q^2}
\frac{\cos(\theta_0+\epsilon)\sin\epsilon}{\sin\theta_0} &
\frac{\cos(\theta_0+\epsilon)\sin(\theta_0+\epsilon)}{\sqrt{6}
\sin\theta_0}\left( \frac{m^2_\omega}{m_\omega^2-q^2}-\frac{m_\phi^2}
{m_\phi^2-q^2}  \right) \\
{\sqrt{6}\cos\epsilon\sin\epsilon\over\sin\theta_0}
\left( \frac{m_\phi^2}
{m_\phi^2-q^2} - \frac{m^2_\omega}{m_\omega^2-q^2}\right)  &
\frac{m_\phi^2}{m_\phi^2-q^2}
{\cos\epsilon\sin(\theta_0+\epsilon)\over\sin\theta_0} -
\frac{m_\omega^2}{m_\omega^2-q^2}{\sin\epsilon\cos(\theta_0
+\epsilon)\over\sin\theta_0}
\end{array} \right)
\nonumber\\*[7.2pt]
& &\times\left(\begin{array}{c}
F_{in}^{I=0}(q^2)\\
F_{in}^s(q^2)
\end{array}\right).
\label{form}
\ee
A couple of instructive features can be directly read off from
this expression. First, the dependence on both the
vector-meson-current and vector-meson-nucleon couplings has dropped
out. This is a straightforward consequence of charge normalization,
which requires both couplings to cancel each other, and is a general
feature of VMD form factors \cite{sa}. As a consequence we have $F(0)
= F_{in}(0)$ for both the strangeness and the isoscalar form factor.
Note that the normalization of the intrinsic isoscalar form factor
[$F_{1, in}^{I=0}(0) = 1/2$] differs from that in Ref.~\cite{nos},
where we used a different isoscalar current. The strangeness radius
and magnetic moment remain the same, however.

Up to now our discussion has been rather general, and different
choices for the intrinsic form factors can be implemented in this
framework. A very important restriction on every such choice is,
however, that it does not lead to double counting with the physics
of the VMD sector. We will come back to this issue later.

In order to calculate the strangeness form factor from
Eq.~(\ref{form}) explicitly, we have to specify the intrinsic
form factors. As in Ref.~\cite{nos}, we adopt the kaon loop model
from the last section \cite{mus93} for the intrinsic strangeness
form factor, but use the physical value for the $\Lambda$ mass
(Ref. \cite{mus93} takes the flavor-symmetric value, {\it i.e.}
the nucleon mass). One could think of adopting an analogous pion
cloud model for the intrinsic isoscalar electromagnetic form
factor. Since, however, typical models for intrinsic nucleon
charge distributions require a large quark core contribution in
addition to the pion cloud \cite{bro}, we do not expect intrinsic
electromagnetic form factors to be sufficiently well modeled by
pion loops, and we will therefore follow a different strategy.

We first adopt H{\"o}hler's fit \cite{hoe76} of the isoscalar
form factor to the experimental data, summarized for convenience
in Table I. From this fit we extract the {\it intrinsic} isoscalar
form factor by inverting the VMD matrix in Eq.~(\ref{form}). Then
we determine the strangeness form factor from the second row of
Eq.~(\ref{form}). The contribution from the intrinsic strangeness
part to the isoscalar form factor is very small and plays almost
no role in the determination of $F_{in}^{I=0}(q^2)$.

Since the strangeness magnetic moment is obtained from the magnetic
form factor at $q^2 = 0$, it is not modified by the vector mesons
and originates solely from the intrinsic contribution. It has
therefore the same value as in the kaon loop model. Both the
Dirac and the Sachs strangeness radii, however, get an additive
contribution from the vector mesons. This contribution increases
the charge radius by about a factor of 3,  $r_{s, Dirac}^2 = -
(0.0243 - 0.0245) \, {\rm fm}^2$, and the Sachs radius by about
a factor of 2: $r_{s}^2 = - (0.040 - 0.045) \,  {\rm fm}^2$. Both
signs are the same as those of the intrinsic contribution.

This enhancement in the strangeness radii is rather important
from the perspective of experiments. The charge radius obtained
from the kaon cloud alone is too small, for example, to be
detected in the planned parity-violating elastic $\vec{e} p$
scattering experiments, whereas the enhancement due to the vector
mesons could lead to an observable effect.

A couple of other aspects of the VMD approach should be noted.
The VMD part of the form factors is generic and independent of
any details of the model except the general VMD form. In
particular, it is independent of the current-meson couplings
with their underlying theoretical assumptions and uncertainties.
It also renders the strangeness radius less sensitive to the
model-dependence of the intrinsic form factors. The kaon cloud
form factors alone are, for example, dominated by the strongly
parametrization-dependent seagull contributions.

Another important aspect of the VMD results is their crucial
dependence on the $\omega$ -- $\phi$ mixing. Indeed, the
strangeness radius would not receive any contribution from ideally
mixed vector meson states, since the nucleon has no overall
intrinsic strangeness. As a consequence, the VMD contribution to
the radius is proportional to the sine of the mixing angle
$\epsilon$.

\subsection{Constituent quark models}
\label{cqmsection}

In this section we will discuss the strangeness radius from the
point of view of a large class of hadron models based on a
constituent quark core. In the ``naive'' nonrelativistic
constituent quark models, the quarks are pointlike and, since sea
quarks are absent, there is no mechanism for a nonvanishing
strangeness distribution in the nucleon. However, it was argued
some time ago \cite{kap88} that the constituent quark picture
can be generalized to accommodate a finite strangeness content.
Indeed, if one regards a constituent quark as a QCD current quark
surrounded by a complicated, nonperturbative cloud of gluons and
$q \qbar$ pairs, then even up and down constituent quarks can
have a strangeness distribution.

Of course, predictions of the nucleon's strangeness distribution
in such constituent quark models have still to relate the
strangeness content of the constituent quark to that of the
nucleon. This is no simple task and generally requires the
solution of a relativistic Fadeev-type equation\footnote{First
studies of this type in the NJL model have recently reproduced
the nucleon mass quite well \cite{hua93}, but they are not easily
generalizable to the calculation of form factors. Furthermore,
it is not obvious that they are consistent with the Hartree-Fock
or large-$N_c$ approximations used to derive the constituent
quark propagators.}.

Fortunately, however, the nucleon's strangeness Sachs radius
can be inferred exactly from that of the constituent quark
without any specific calculation, and it is independent of the
inter-quark dynamics. To see this, consider a system of three
constituent quarks with strange charge distributions
$\rho_s(\vec{r} - \vec{r}_i)$ of identical shape, centered
at (in general time-dependent) positions $\vec{r}_i$. The
total strangeness radius of the nucleon is then
\be
r^2_s = \int d V \, \vec{r \,}^2 \left[ \, \rho_s(\vec{r} -
\vec{r}_1) + \rho_s(\vec{r} - \vec{r}_2) + \rho_s(\vec{r} -
\vec{r}_3) \right],
\ee
and after shifting the integration variable by the individual
positions of the quarks one obtains
\be
r^2_s &=& 3 \int d V \, \vec{r}^{\, 2} \rho_s(\vec{r}) \, +
\, 2 \,(\vec{r}_1 + \vec{r}_2 + \vec{r}_3) \cdot \int d V \,
\vec{r} \, \rho_s(\vec{r}) \, + \, (\vec{r}_1^{\, 2} +
\vec{r}_2^{\, 2} + \vec{r}_3^{\, 2}) \int d V \, \rho_s(\vec{r})
\nonumber \\ &=& 3 \, (r^2_s)_q.
\ee
The last equation holds for isotropic quark strangeness
distributions with vanishing overall strangeness charge and
shows that the nucleon strangeness radius is, under the stated
conditions, just the sum of the quark strangeness
radii\footnote{Note that the isotropy condition could be
violated by the strangeness current (as opposed to charge)
distribution, which prevents us from extending the above
argument to the strange magnetic moment. Note also that
two- or three-body correlations between the quarks, beyond
or instead of the common mean field potential, could invalidate
the isotropy condition. }.

The above ideas can be studied quantitatively in models which
generate constituent quarks dynamically by dressing the
elementary quarks with $q \qbar$ pairs. The prototype of this
class of models has been introduced by  Nambu and Jona-Lasinio
(NJL)\cite{nam61}, and the first study of the scalar strangeness
content of constituent quarks in this framework can be found
in Ref. \cite{ber88}.

In order to give an estimate of the nucleon strangeness radius
in constituent quark models, all what remains to be done is to
calculate the strangeness radius of the constituent quark itself.
We will perform such a calculation in the remainder of this
section in the context of the NJL model. An extensive study of
electromagnetic quark and meson properties in different $SU(3)$
generalizations of the NJL model can be found in
Refs.~\cite{we1,we2}, and we will employ some of their
results.

A strangeness component in the valence constituent quarks can
only be generated by OZI-rule violating processes, which require
a flavor mixing interaction. In the conventional Hartree-Fock
approximation to the NJL model (on which the work of Ref.
\cite{we2} is based), flavor mixing originates exclusively from
determinantal six-quark interactions \cite{ber88}. These terms,
of the form of 't Hooft's instanton-generated effective lagrangian
\cite{tho}, represent the anomalous breaking of the $U(1)_A$
symmetry in QCD, and contain two coupling constants $H$ and
$H^\prime$. Their physical role is primarily to generate the mass
splittings of the singlet and octet states in the pseudoscalar
channel ($H$) and of the $\rho$ and $\omega$ mesons in the vector
channel ($H^\prime$). In addition, these couplings will determine
the strangeness content of the constituent quark.

We start our calculation by writing the constituent $U$ quark
matrix elements of the strangeness current as
\be
\bra{ U(p^\prime)}J_\mu^s(0)\ket{U(p)}&
=&\overline{U}(p^\prime)\left(\gamma_\mu f_{1}^s(q^2) +
 i\frac{\sigma_{\mu\nu}q^\nu} {2m_u} f_{2}^s(q^2)
\right)U(p)\; \nonumber\\*[7.2pt] & =& \sum_\nu \,
[{\cal F}(q^2)]_
{\mu,\nu}^{s,u}\overline{U}(p^\prime)\Gamma^\nu U(p).
\label{matel}
\ee
The form factor matrix ${\cal F}$ is, in Hartree-Fock
approximation, a solution of the Bethe-Salpeter type
equation
\be
{\cal F} = 1 + {\cal KJF}\; ,
\ee
where ${\cal K}$ is the effective two-body reduction of
the NJL interaction, containing the coupling constants,
and ${\cal J}$ is the so-called generating correlation
function, which is basically a two-body propagator. The
explicit forms of both ${\cal K}$ and ${\cal J}$ are given
in Ref.~\cite{we1}.

The Sachs strangeness radius of the constituent $u$
quark,
\be
(r^2_s)_u = \left.6\frac{ d g_E^s(q^2)}{ d q^2}\right|_
{q^2=0} \; ,
\ee
can now be obtained from the explicit solutions for $f_1^s$
and $f_2^s$, given in \cite{we2}, and $g_E^s = f_1^s +
(q^2 / 4m_u^2) f_2^s$. We find
\be
(r^2_s)_u = \frac{96 \sqrt{2}
H^\prime <\ubar u>}{D^{I=0}(0)}m_u m_sI_s
\left(-2G_2I_u + \frac{1}{4m_u^2}\right) \; ,
\label{ra}
\ee
where
\be
D^{I=0}(0)=1-8m_u^2 H^\prime <\sbar s> I_u-128 m_u^2 m_s^2
(H^\prime <\ubar u>)^2 I_u I_s
\ee
and the integrals $I_q$ can be solved analytically:
\be
I_q = \frac{3}{4\pi^2}\left[\frac{\Lambda^2}{\Lambda^2+m_q^2} +
\log\left(\frac{m_q^2}{\Lambda^2+m_q^2}\right)\right] \; .
\ee
With the parameters of Ref.~\cite{we1}, $m_u=m_d=364\MeV$,
$\, m_s=522\MeV$, $\, \Lambda=0.9\GeV$, $\, G_2 \Lambda^2 =
2.51$ and $H^\prime \Lambda^2 <\sbar s> = -4.4 \times10^{-2}$,
we finally obtain the numerical value
\be
r^2_s = 3 \, (r^2_s)_u = 1.69\times 10^{-2}\fm^2 .
\ee
It comes somewhat as a surprise that $r^2_s$  has the opposite
sign of the VMD result, since the isoscalar and isovector form
factors of the NJL constituent quarks are also dominated by
vector meson intermediate states \cite{we2}. However, Eq.
(\ref{ra}) shows that the NJL result is proportional to
$H^\prime$ (which sets the singlet-octet mass splitting in
the vector sector), whereas $\omega-\phi$ mixing, which gave
rise to the VMD contribution to $r^2_s$, occurs also for
$H^\prime=0$ (in the case of $m_u^0\neq m_s^0$). This indicates
that physics other than VMD contributes substantially to the NJL
result. Note furthermore that the entire result depends on
subleading effects in 1/$N_c$ counting.

In Table II we summarize all the theoretical estimates of the
nucleon strangeness radius discussed above.

\section{The strangeness form factors}

\label{vmd}

Our discussion has so far mainly focused on the first
nonvanishing moments of the nucleon strangeness distribution.
These quantities -- the strangeness radius and magnetic moment
-- will be the first characteristics of the strange form factors
to be measured, and thus have received most of the theoretical
attention.

However, not only does the higher momentum region of the form
factors contain important physical information, but its knowledge
is also required for the interpretation of the currently
prepared experiments to measure $r^2_s$ and $\mu_s$ (and the
already discussed BNL neutrino scattering experiment). Since
the form factors cannot be measured exactly at $q^2 =0 $, the
data have to be extrapolated back to the light point from small
spacelike $q^2$. The extracted values for $r^2_s$ and $\mu_s$
will thus depend on the assumed momentum dependence of the form
factors, and its better understanding is clearly important.

In the absence of reliable information one sometimes parametrizes
the momentum dependence, as in the electromagnetic form factors,
in Galster's dipole form. This choice is motivated by simplicity
and convenience, but has no theoretical basis. The physics
contributing to the strange form factors -- originating exclusively
from sea quarks -- might well lead to a momentum dependence other
than the dipole form (which also does not reproduce the asymptotic
behavior suggested by quark counting rules). Note, however, that
a Galster form with independent mass parameters to be determined
by experiment, as proposed in Ref. \cite{mus92}, could be
sufficiently flexible to describe the form factors in the momentum
range relevant for CEBAF.

In the present section we will extend our work on the
simple VMD plus kaon cloud model (proposed in Ref. \cite{nos} and
discussed in the last section) to the calculation of the strangeness
vector form factors in the momentum region accessible at CEBAF.
Since our approach is based on low-energy dynamics, however, it
will not be applicable at larger $q^2$.

The momentum dependence of the VMD contribution was already given
in Eq. (\ref{form}). The intrinsic form factors will be determined
in the kaon cloud model of Ref.~\cite{mus93}. Even if the
corresponding kaon loop graphs to be calculated are U.V. finite,
the effective hadronic description of the underlying physics
breaks down at large momenta. We will therefore follow Ref.
\cite{mus93} and attach form factors,
\be
H(k^2) = \frac{m_K^2 - \Lambda^2}{k^2 - \Lambda^2} \; ,
\label{fa}
\ee
taken from the Bonn potential \cite{holz}, to the meson-nucleon
vertices to cut off the loop momenta. The Bonn values for the
cutoff $\Lambda$ in the $N\Lambda K$ vertex were extracted from
fits to baryon-baryon scattering data and lie in the range of
1.2 -- 1.4 $\GeV$.

As already mentioned, the $J^s_\mu$ - hadron couplings will be
assumed pointlike and therefore given by the strangeness charge
of the struck hadron. Also, in order to maintain gauge invariance
in the presence of the extended meson--baryon vertices, we have
to introduce seagull vertices. These vertices are generated from
the effective lagrangian
\be
{\cal L}_{N\Lambda K} =-ig_{N\Lambda K} \bar{\Psi}\gamma_5 \Psi
H(-\partial^2) \phi \; ,
\ee
where $\Psi$ and $\phi$ represent baryon and kaon fields, by minimal
substitution in the gradient operator. This prescription \cite{ohta},
while not unique, generates the seagull vertex
\be
i\Gamma_\mu(k,q)=\mp g_{N\Lambda K} Q_K \gamma_5(q\pm 2k)_\mu
\frac{H(k^2) - H((q\pm k)^2)}{(q\pm k)^2 - k^2} \; ,
\label{seagull}
\ee
where $Q_K$ is the kaon strangeness charge, and the upper (lower)
sign corresponds to an incoming (outgoing) meson. As expected,
this vertex disappears in the limit $\Lambda\rightarrow\infty$ of
pointlike couplings.

We can now identify three distinct contributions to the intrinsic
form factors. The three corresponding amplitudes are associated
with processes in which the current couples either to the baryon
line (B), the meson line (M) or the meson-baryon vertex (V) in
the loop, and are given by
\be
\Gamma^B_\mu(p^\prime,p) = -ig^2_{N\Lambda K} Q_\Lambda \int
\frac{d^4k}{(2\pi)^4} \Delta(k^2) H(k^2) \gamma_5 S(p^\prime - k)
\gamma_\mu S(p - k) \gamma_5 H(k^2) \; ,
\label{bv}
\ee
\be
\Gamma^M_\mu(p^\prime,p) = -ig^2_{N\Lambda K} Q_K \int
\frac{d^4k}{(2\pi)^4} \Delta((k+q)^2) (2k+q)_\mu \Delta(k^2)
H((k+q)^2) \gamma_5 S(p - k)  \gamma_5 H(k^2) \; ,
\label{mv}
\ee
\be
\Gamma^V_\mu(p^\prime,p)& =& -ig^2_{N\Lambda K} Q_K \int \frac{d^4k}
{(2\pi)^4} H(k^2) \Delta(k^2) \left[\frac{ (q+2k)_\mu}{ (q+k)^2-k^2}
\left(H(k^2)\, - H((k+q)^2)\right) \times \right.
\nonumber\\*[7.2pt]
& &\left.  \gamma_5 S(p-k) \gamma_5 - \frac{ (q-2k)_\mu}{
(q-k)^2-k^2} \left(H(k^2)-H((k-q)^2)\right) \gamma_5 S(p^\prime-k)
\gamma_5\right] \; .
\label{vv}
\ee
$\Delta(k^2) = (k^2-m_K^2+i\epsilon)^{-1}$ is the kaon
propagator, $S(p - k) = (\rlap{/}{p}-\rlap{/}{k}-
M_\Lambda+i\epsilon)^{-1}$
is the $\Lambda$ propagator, $p^\prime=p+q$ with $q$ being
the photon momentum, and $Q_\Lambda$ is the $\Lambda$
strangeness charge. In our convention (where the s-quark has
strangeness $+1$) $Q_\Lambda=1$ and $Q_K=-1$.
The values of the coupling and masses used are $M_N=939\MeV$,
$M_\Lambda=1116\MeV$, $m_K=496\MeV$ and $g_{N\Lambda K}/
\sqrt{4\pi}=-3.944$ \cite{holz}. It is easy to show that these
three amplitudes satisfy the Ward-Takahashi identity.

The intrinsic strangeness form factors are obtained by writing
the nucleon matrix element of the sum of these amplitudes in
terms of Dirac and Pauli form factors
\be
\overline{N}(p^\prime)
\Gamma_\mu(p^\prime,p)N(p)
= \overline{N}(p^\prime)\left(\gamma_\mu F_{1,in}^s(q^2) +
 i\frac{\sigma_{\mu\nu}q^\nu} {2M_N} F_{2,in}^s(q^2)
\right)N(p) \; ,
\ee
where $N$ is the nucleon spinor. The corresponding electric and
magnetic Sachs form factors can be obtained from Eq. (\ref{sachs}).

The dependence of the electric strangeness form factor on the
(space-like)  momentum transfer $Q^2= - q^2$ is shown as the full
line in Fig. 1 and, for comparison, the result of the pole fit
model \cite{ja} is also plotted. The results are quite different
both in size and magnitude, and document the present state of
theoretical uncertainty. We further plot the intrinsic contribution
alone, which essentially corresponds to the full form factor in
the model of Ref. \cite{mus93}. While it has the same sign as
our result and is of comparable magnitude, its slope at the
origin, and consequently the mean square strangeness radius, is
by a factor of 3 smaller.

It is interesting to see how much of the form factor in our
approach comes from the VMD contribution alone, i.e. by treating
the core as pointlike and by setting  $F_{1, in}^{I=0}(Q^2)=1/2$
and $F_{1, in}^s (Q^2)=0$. While the slope at the origin is
slightly reduced in this case, the overall behavior of the form
factor remains almost unchanged.

In Fig. 2 we show the dependence of $G^s_E$ on the parameters
of the intrinsic form factors. While H{\"o}hler's three different
fits for the intrinsic isoscalar form factor $F_{in}^{I=0}$ have
almost no influence on our result and cannot be distinguished
in Fig. 2, we see a mild dependence on the kaon loop cutoff in
the range between $\Lambda=1.2\GeV$ and  $\Lambda=1.4\GeV$, which
is compatible with the Bonn potential fits. We also plot the pole
fit result \cite{ja} for the three fits (fit nos. 7.1, 8.1 and 8.2)
of the isoscalar form factor and find a rather strong dependence.

Figure 3 shows the magnetic strangeness form factor in our model
and again, for comparison, the pole fit result. As in the case of
the electric form factor, their slopes at the origin, their
curvatures and their momentum dependence are generally rather
different. However, the values at $Q^2 = 0$, {\it i.e.} the strange
magnetic moments, are quite similar. Since the VMD contribution
does not alter the value of $\mu_s$, given by $G_{M,in}^s (0)$,
the intrinsic form factor starts at the same value as the full
$G_M^s$ but is, in contrast to $G_E^s$, considerably smaller. The VMD
contribution alone, however, is not as close to the complete result
as in the case of $G_E^s$. (To generate this graph we kept the
intrinsic isoscalar and strange form factors constant, at their
values at $Q^2 = 0$.)

The G0 experiment at CEBAF \cite{ce17} will measure the $Q^2$
dependence of $G_M^s$ in the momentum range $0.1 \GeV^2<Q^2< 0.5
\GeV^2$, with a resolution $\delta\mu_s \simeq \pm0.22$ at low
$Q^2$, and decreasing for larger $Q^2$. The SAMPLE experiment at
MIT-Bates \cite{mit} will determine $\mu_s$ with comparable
experimental error, and the combined data might thus be sufficient
to distinguish between our and the pole result.

The dependence of $G_M^s$ on the cutoff and on the choice of
fit for $F_{in}^{I=0}$ can be seen in Fig. 4. While our result
is still practically insensitive to the choice of fit, the
variation with the kaon loop cutoff, again in the range
$\Lambda= (1.2 - 1.4) \GeV$, is stronger than for the electric
form factor. The dependence of the pole fit result on the choice
of fit is even larger, as for $G_E^s$.

We have already mentioned that some authors, and in particular
Garvey {\it et al.} in their reanalysis of the BNL neutrino
scattering experiment \cite{gar93}, used a Galster dipole
parametrization
\be
F_1^s(Q^2) = {F_1^s Q^2\over \left(1+{Q^2\over4M_N^2}\right)
\left(1+{Q^2\over M_V^2}\right)^2} \; ,
\label{f1}
\ee
\be
F_2^s(Q^2) = {F_2^s(0)\over
\left(1+{Q^2\over M_V^2}\right)^2}\; ,
\label{f2}
\ee
for the $Q^2$ dependence of the strange form factors, in which
$M_V=0.843\GeV$ is taken to be the same as in the electromagnetic
form factors. As mentioned before, a more general form of the
Galster parametrization, in which the masses are taken as
independent fit parameters, has been proposed in Ref. \cite{mus92}.
The reanalysis of the E734 experiment gave a first idea of the
leading momentum dependence, $F_2^s(0)=-0.40\pm0.72$ and $F_1^s
= -(1/6) (r_s^2)_{Dirac} = (0.53\pm0.70)\GeV^2$, although these
values have large error bars and are consistent with zero.
Furthermore, the assumptions which had to be made for the neutron
form factors add to the theoretical uncertainties of this reanalysis.

Still, the estimates of Ref. \cite{gar93} allow us to assess the
compatibility of the dipole form with our results and those of
the pole fit. In Figs. 5 and 6 we compare our Dirac and Pauli
form factors and those of Ref. \cite{ja} with the dipole forms,
Eqs.(\ref{f1}) and (\ref{f2}). The differences between the three
approaches are clearly more pronounced for $F_1^s$. Figure 5 also
indicates that the value for the Dirac strangeness radius, if
extracted by fitting the momentum dependence of the dipole
parametrization, can be rather different  from the one obtained
in other estimates. The same holds for the strange magnetic
moment, as can be seen in Fig. 6. On the other hand, all three
results for $F_2^s(Q^2)$ (Fig. 6) look rather similar in the
momentum region $0.4 \GeV^2<Q^2< 0.8 \GeV^2$, in which the data
will be taken at CEBAF.

Finally, comparing our approach to some of the other proposed
mechanisms, we note interesting differences in the role played
by vector meson mixing. Whereas the rather small deviation of
the $\omega$ and $\phi$ states from ideal mixing is indispensable
for a non-vanishing VMD contribution to the strangeness radius
in our approach, this is not so in the pole fit model of Ref.
\cite{ja}.

Indeed, in the pole model $r^2_s$ gets bigger as the mixing
angle $\epsilon$ goes to zero. In our approach, on the other
hand, the Dirac form factor reduces in the same limit to
\be
F^s(q^2) = {m_\phi^2\over m_\phi^2-q^2} F_{in}^s(q^2) \; ,
\ee
and the strangeness current interacts with the nucleon only
through the $\phi$ meson. In particular, the strangeness radius
does not get any contribution from VMD in the limit of ideal
mixing since the nucleon has no intrinsic strangeness
[{\it i.e.} $F_{in}^s(0) = 0$].

\section{Summary and Conclusions}
\label{5section}

In this paper we present a theoretical estimate for the
strangeness
vector form factors of the nucleon in the momentum region
relevant for the planned experiments at MIT-Bates and CEBAF.
This estimate is based on our previously introduced model of
the nucleon strangeness distribution in terms of a kaon cloud,
a vector meson dominance contribution  and $\omega$ - $\phi$
mixing.

We find both Sachs form factors to be dominated by the VMD
contribution. The electric strangeness form factor, in
particular, is left practically unchanged if the intrinsic
contribution is taken point-like. For the same reason
the dependence of our result on the meson-baryon form factor
masses in the kaon loops is very weak. Both the sign and the
slope of the electric form factor are opposite to those of the
pole-fit estimate. Together with the considerably larger slope
of the pole-fit result at zero momentum transfer this leads
already at moderate momentum transfer to a significant
difference in magnitude between the two form factors.

Although the intrinsic kaon cloud contribution plays a somewhat
more important role in the magnetic Sachs form factor, its overall
shape and slope are again mainly determined by the vector mesons.
The value at zero momentum transfer, {\it i.e.} the strangeness
magnetic moment, however, is not modified by the VMD contribution
and thus shows a more pronounced dependence on the intrinsic
meson-baryon form factors. As in the case of the electric form
factor, the overall qualitative behavior of our result, and in
particular its slope and curvature, differs considerably from
the pole fit estimate.

We also compare our results to a Galster dipole parametrization,
with the cut-off masses fixed at the same values as in the
electromagnetic form factors, which has been used to reanalyze
the BNL E734 neutrino scattering experiment. The Dirac dipole
form factor has the same sign and curvature as ours, but with a
considerably larger overall size and slope at the origin,
whereas the Pauli form factors agree rather closely for the
momenta to be probed at CEBAF, {\it i.e. } for $Q^2 > 0.4
\GeV^2$.

Since our results, with the exception of the strange magnetic
moment, are dominated by the vector meson sector, they show a
largely reduced sensitivity to the rather strong model
dependence of the intrinsic contribution. The intrinsic form
factors from the kaon cloud model adopted in this paper, for
example, receive their main contribution from the seagull
vertices, which are determined by a minimal, but not unique
prescription.

The vector meson dominance mechanism, on the other hand, has a
robust and largely generic character. It introduces no free
parameters and does not leave much freedom in its implementation.
In particular, however, it is based on  phenomenologically
successful physics which is well established from the
electromagnetic interactions of hadrons.

We further study the nucleon strangeness radius from the point
of view of a class of constituent quark models in which the
constituent quarks have an extended structure.  The valence up
and down quarks in the nucleon can then acquire an intrinsic
strangeness distribution. We point out that the strangeness
radius of the nucleon is model-independently given by the sum
of the constituent quark radii, since their strangeness charge
distribution is isotropic and since their overall strangeness
is zero. For a quantitative estimate we employ the
Nambu-Jona-Lasinio model and find a 2--3 times smaller value
than in the kaon-loop-VMD model with the opposite sign.

In order to put these results into perspective, we also review
and discuss some of the other existing theoretical estimates.
Comparing the results for the strangeness radius and magnetic
moment reveals in particular the large discrepancies between
those predictions. Both signs of the strangeness radius, for
example, and values within a range of an order of magnitude
have been found. Clearly the theoretical uncertainties are at
present not reliably under control and the existing results
have to be regarded as order-of-magnitude estimates.

This situation reflects both the size of the challenge with
which the non-valence strangeness sector confronts existing
hadron models, and our present lack of insight from the first
principles of QCD. While first attempts to study the effects of
disconnected quark loops on the lattice have been reported
\cite{keh-fei}, it will still take considerable time and effort
before reliable results for the strange quark content of the
nucleon can be expected. In the meantime, however, the data
from Bates and CEBAF will constrain the existing hadron models
and test their underlying physics. It thus seems guaranteed
that the subject will remain intriguing and challenging in
the years to come.

\acknowledgements

T.D.C., H.F. and X.J. acknowledge support from the Department
of Energy
under Grant No.\ DE--FG02--93ER--40762.
T.D.C. and X.J. acknowledge additional support from the
National Science
Foundation under Grant No.\ PHY--9058487.
M.N. acknowledges the warm hospitality and congenial atmosphere
provided by the University of Maryland Nuclear Theory Group and
support from FAPESP  BRAZIL.


\newpage

\begin{table}
\caption{Fit parameters from Ref.~\protect\cite{hoe76}.
$F_1^{I=0}(q^2) =
\sum_{i} a_i/(m_i^2-q^2)$, $ F_2^{I=0}(q^2) =
\sum_{i} b_i/(m_i^2-q^2)$. }
\begin{tabular}{|ccccc|}
fit number &  & & & \\
\hline
8.1 & $m_V(\GeV)$ & 0.78 & 1.02 & 1.40\\
 & $a_i(\GeV)$ & 0.71  & -0.64 & -0.13\\
 & $b_i(\GeV)$ & -0.11 & 0.13 & -0.02\\
\hline
8.2 & $m_V(\GeV)$ & 0.78 & 1.02 & 1.80\\
 & $a_i(\GeV)$ & 0.69  & -0.54 & -0.21\\
 & $b_i(\GeV)$ & -0.14 & 0.20 & -0.07\\
\hline
7.1 & $m_V(\GeV)$ & 0.78 & 1.02 & 1.67\\
 & $a_i(\GeV)$ & 0.68  & -0.55 & -0.24\\
 & $b_i(\GeV)$ & -0.16 & 0.25 & -0.08\\
\end{tabular}
\label{tab-1}
\end{table}

\begin{table}
\caption{Theoretical results for the strange magnetic
moment and strangeness radius}
\begin{tabular}{|c|c|c|c|}
 source & $\mu_s (\mu_N) $ & $r^2_s (\fm^2)$ & Ref. \\
\hline
poles  & $-0.31 \pm 0.09$ & 0.14 $\pm$ 0.07 & \cite{ja}\\
 kaon-loops (SU(3)-symm.) & - (0.31 -- 0.40)  &
- (2.71 -- 3.23)x$10^{-2}$ & \cite{mus93}\\
 kaon-loops & - (0.24 -- 0.32)   &
-(2.23 -- 2.76)x$10^{-2}$ & \cite{nos}\\
VMD & - (0.24 -- 0.32)  &  -(3.99 --
4.51)x$10^{-2}$ & \cite{nos} \\
NJL & & 1.69x$10^{-2}$ & this work \\
SU(3) Skyrme & -0.13  & -0.11 & \cite{par91} \\
\end{tabular}
\label{tab-2}
\end{table}

\newpage

\begin{figure}
\caption{The $Q^2$ dependence of the electric strangeness form
factor
using $\Lambda=1.2\GeV$.
The solid line represents our result. The long-dashed line
gives the
intrinsic contribution and the short-dashed line gives only
the VMD
contribution. The dotted line is the result of
Ref.~\protect\cite{ja} with
the fit 8.1 of Table I.}
\label{fig-1}
\end{figure}
\begin{figure}
\caption{Same as Fig. 1 but for two values of $\Lambda$.
The upper solid
and long-dashed lines are for $\Lambda=1.4\GeV$ and the
lower solid
and long-dashed lines are for $\Lambda=1.2\GeV$. The dotted
lines
are the results of  Ref.~\protect\cite{ja} for the three
different fits in
Table I.}
\label{fig-2}
\end{figure}
\begin{figure}
\caption{The $Q^2$ dependence of the magnetic strangeness
form factor
using $\Lambda=1.2\GeV$.
The solid line represents our result. The long-dashed line
gives the
intrinsic contribution and the short-dashed line gives only
the VMD
contribution. The dotted line is the result of
Ref.~\protect\cite{ja} with
the fit 8.1 of Table I.}
\label{fig-3}
\end{figure}
\begin{figure}
\caption{Same as Fig. 3 but for two values of $\Lambda$. The
upper solid
and long-dashed lines are for $\Lambda=1.2\GeV$ and the lower
solid
and long-dashed lines are for $\Lambda=1.4\GeV$. The dotted
lines
are the results of  Ref.~\protect\cite{ja} for the three
different fits in
Table I.}
\label{fig-4}
\end{figure}
\begin{figure}
\caption{The $Q^2$ dependence of the Dirac strangeness form
factor
using $\Lambda=1.2\GeV$.
The solid line represents our result. The dashed and dotted
lines
are the results of Refs.~\protect\cite{gar93} and
\protect\cite{ja} respectively.}
\label{fig-5}
\end{figure}
\begin{figure}
\caption{Same as Fig. 5 for the Pauli strangeness form factor.
Note the similarity of all three predictions at intermediate
momentum transfers.}
\label{fig-6}
\end{figure}

\end{document}